\documentclass{article}
\usepackage{spconf,amsmath,graphicx}
\usepackage{makecell}

\title{Frequency and temporal convolutional attention for text-independent speaker recognition}
\name{Sarthak Yadav, Atul Rai\thanks{This research is funded by Staqu Technologies, India.}}
\address{Staqu Technologies, India}
\begin{document}
%
\maketitle
\begin{abstract}
Majority of the recent approaches for text-independent speaker recognition apply attention or similar techniques for aggregation of frame-level feature descriptors generated by a deep neural network (DNN) front-end. In this paper, we propose methods of convolutional attention for independently modelling temporal and frequency information in a convolutional neural network (CNN) based front-end. Our system utilizes convolutional block attention modules (CBAMs) \cite{woo2018cbam} appropriately modified to accommodate spectrogram inputs. The proposed CNN front-end fitted with the proposed convolutional attention modules outperform the no-attention and spatial-CBAM baselines by a significant margin on the VoxCeleb \cite{nagrani2017voxceleb,chung2018voxceleb2} speaker verification benchmark. Our best model achieves an equal error rate of $2.031\%$ on the VoxCeleb1 test set, which is a considerable improvement over comparable state of the art results. 
For a more thorough assessment of the effects of frequency and temporal attention in real-world conditions, we conduct ablation experiments by randomly dropping frequency bins and temporal frames from the input spectrograms, concluding that instead of modelling either of the entities, simultaneously modelling temporal and frequency attention translates to better real-world performance.
\end{abstract}
\section{Introduction}
\label{sec:intro}

Majority of the recent strides in the field of text-independent speaker recognition can be ascribed to deep neural network (DNN) based speaker embeddings, which have far surpassed conventional state-of-the-art systems such as the i-vector+PLDA framework.

End-to-end deep learning-based speaker recognition systems usually comprise of two components: (i) DNN front-end for extraction of frame-level features; and (ii) temporal aggregation of these frame-level features to an utterance-level embedding.
Majority of the recent works utilize convolutional neural network (CNN) based front-end models for extracting frame-level feature descriptors from spectrogram inputs.

Although sub-optimal since it does not differentiate between frames on the basis of content, temporal averaging is amongst the most frequently used techniques for aggregation of frame-level features \cite{nagrani2017voxceleb, chung2018voxceleb2, Yadav2018}. A number of recent works have proposed the use of statistical or dictionary based methods for aggregation to mitigate this problem. \cite{snyder2017deep} proposed the statistics pooling layer, which combines mean and standard deviation statistics for weighted aggregation of temporal frames.
More recently, \cite{hajavi2019deep} proposed time-distributed voting (TDV) for aggregating features extracted by their UtterIdNet front-end in short segment speaker verification, especially sub-second durations.
\cite{xie2019utterance} proposed the usage of dictionary-based NetVLAD or GhostVLAD \cite{zhong2018ghostvlad} for aggregating temporal features, using a 34-layer ResNet based front-end for feature extraction. Numerous recent works \cite{bhattacharya2017deep, Zhu2018, okabe2018attentive, India2019} have proposed attention based techniques for aggregation of frame-level feature descriptors, to assign greater importance to the more discriminative frames. 

A prominent attention mechanism in the domain of computer vision is convolutional attention \cite{woo2018cbam, wang2017residual},
which facilitates modelling of spatial and channel attention throughout the entire CNN feature extraction network. In this paper, we propose methods of convolutional attention based on \textit{Convolutional block attention module} (CBAM) \cite{woo2018cbam} for speaker verification.
The main contributions of this work are two-fold: (i) We propose convolutional attention modules based on CBAM for modelling frequency and temporal attention, \textit{viz.} \textit{f}-CBAM and \textit{t}-CBAM, along with an equal-weighted composite module for capturing both frequency and temporal attention, called \textit{ft}-CBAM; and, (ii) We conduct ablation experiments for a more thorough assessment of the proposed attention modules as well as their performances under real-world conditions, concluding that instead of modelling either of the entities, simultaneously modelling temporal and frequency attention translates to better real-world performance.

\section{Related works}
\label{sec:related}

Attention mechanisms have led to significant advances across computer vision, spoken language understanding and natural language processing,
increasing the modelling capacity of deep neural networks by concentrating on crucial features and suppressing unimportant ones. For speaker recognition, \cite{bhattacharya2017deep, Zhu2018} utilize self-attention for aggregating frame-level features. \cite{okabe2018attentive} combined attention mechanism with statistics pooling \cite{snyder2017deep} to propose attentive-statistics pooling. Most recently, \cite{India2019} employ the idea of multi-head attention 
\cite{vaswani2017attention} for feature aggregation, outperforming an I-vector+PLDA baseline by 58\% (relative). 
However, by applying attention or similar techniques only on the feature descriptors generated by the DNN front-end and not \textit{throughout} the front-end model, majority of the recent works are (i) not fully utilising the representation power of DNN front-end models; and (ii) implicitly modelling temporal attention alone in the process. As opposed to the methods mentioned above, the proposed modules apply attention in the feature extraction module, innately improving the representation capabilities of the model.

Recently, \cite{You2019} proposed the usage of Gated Convolutional Neural Networks (GCNN) for speaker recognition. Matched with a gated-attention pooling method for frame-level feature aggregation, they evaluate the performance of GCNN in an x-vector \cite{snyder2018x} system on SRE16 and SRE18 datasets. In comparison, we propose add-on modules that explicitly model frequency and temporal attention. \cite{xu2019spatial} proposed an encoder-decoder style attention module similar to \cite{wang2017residual}, for extracting spatial and channel attention for automatic speech recognition in noisy conditions. In contrast, we propose convolutional attention modules based on \cite{woo2018cbam} that model frequency and temporal attention along with channel attention, which drastically outperform spatial attention baseline for speaker verification.

\subsection{CBAM: A brief overview}
\label{ssec:cbamover}
Recently, \cite{woo2018cbam} proposed a new network module, named "Convolutional block attention module" (CBAM), which sequentially applies channel attention and spatial attention submodules on the input feature maps. 

CBAM comprises of two components, \textit{viz.} the \textit{channel attention module} and the \textit{spatial attention module}. The following equations can be used to summarize the overall attention process:
\begin{equation}
    F^{'} = M_{c}(F)\bigotimes F
    \label{equation:eq1}
\end{equation}
\begin{equation}
    F^{''} = M_{s}(F^{'}) \bigotimes F^{'},
    \label{equation:eq2}
\end{equation}
where $\bigotimes$ denotes element-wise multiplication, $F$ is the input feature map, $F^{''}$ is the final output of the CBAM module, and $M_{c}$ and $M_{s}$ denote the channel and spatial attention operations, respectively. 
The channel attention module exploits inter-channel relationship of features and generates a 1-D channel attention map by squeezing the spatial dimensions of the input feature map by max-pooling and average pooling, followed by projection using a shared MLP layer. The spatial attention module utilizes the inter-spatial relationship of features, focusing on the spatial location of objects of interest. It applies and concatenates outputs of average-pooling and max-pooling operations along the channel axis which generates an efficient feature descriptor, followed by a 7x7 convolution layer. 

However, unlike computer vision where the modality represents highly correlated points in space and the axes represent the spatial location of an object in a cartesian coordinate system, the axes of a spectrogram represent entirely different domains: frequency and time. This disconnect between the entities represented by the axes of the feature space of the two modalities necessitates the need for targeted convolutional attention modules since preconditions required by existing methods of convolutional attention to effectively model attention in the speech domain might no longer apply.

\section{Proposed Approach}
\label{sec:proposed}
The channel attention module (Eq.\ref{equation:eq1}) extracts general information regarding channel importance in the input feature map, and is used as is.
We propose appropriate changes to the spatial attention submodule for modelling frequency and temporal attention, \textit{viz.} \textit{f}-CBAM and \textit{t}-CBAM, respectively, for spectrogram inputs.

Hence, the input to our proposed modules is $F^{'}$ (Eq. \ref{equation:eq1}), such that $F^{'} \in \mathcal{R}^{C \times H \times T}$  where $C$ denotes the number of input channels, and $H$ and $T$ denote the dimensions along the frequency and temporal axes, respectively.
\subsection{\textit{f}-CBAM}
\label{ssec:fcbam}
For modelling frequency attention, we need to limit the receptive field of the attention module to focus only on the y-axis of the input.

\begin{figure}[htb]
\begin{minipage}[b]{1.0\linewidth}
  \centering
    \includegraphics[width=9cm]{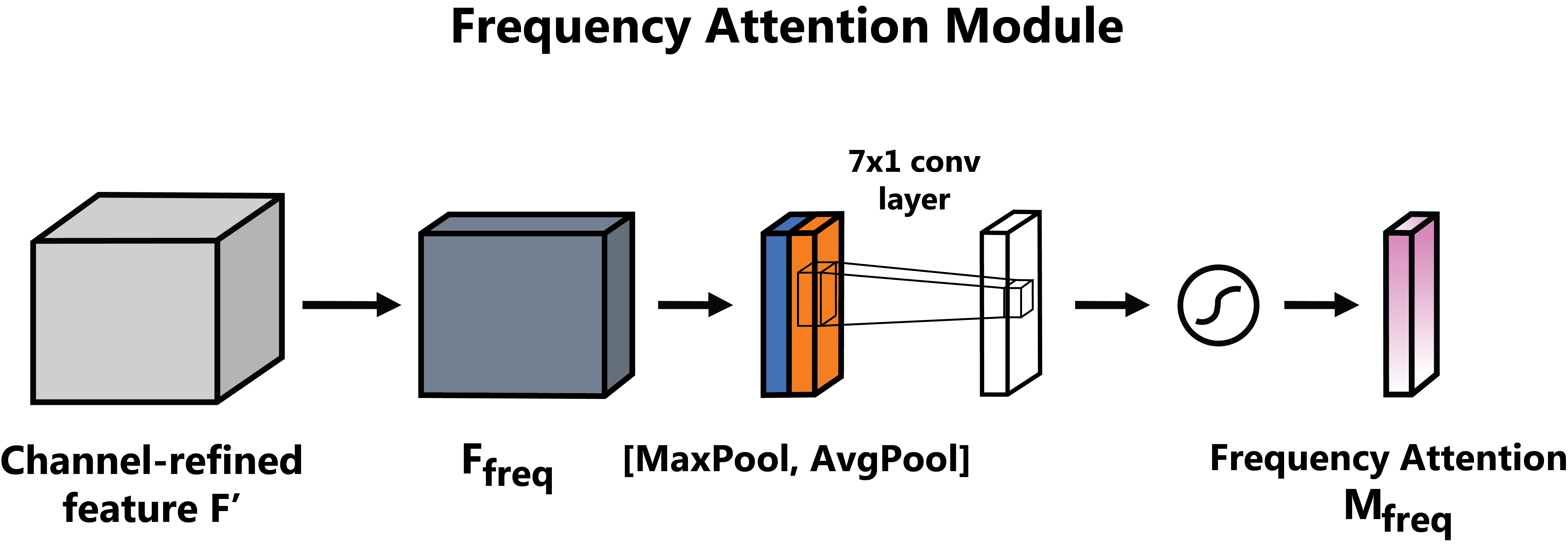}
\end{minipage}
\caption{Proposed \textit{f}-CBAM Module. z-axis represents temporal axis (pictorially represented using dimensions $>1$).}
\label{fig:fcbamfig}
\end{figure}

We aggregate temporal information averaging the input feature map $F^{'}$ along the x-axis to generate an efficient feature descriptor $F_{freq} \in \mathcal{R}^{C \times H \times 1} $ which essentially assigns equal statistical importance to each temporal frame.
\begin{equation}
    F_{freq} = AvgPool_{1 \times T} (F^{'})
    \label{equation:avgpoolt}
\end{equation}
where $AvgPool_{1 \times t}$ represents the average pooling operation with a kernel of size $1\times T$ over the input feature map.

Similar to spatial attention submodule, we then aggregate channel information by generating two feature maps: ${F_{avg}^{f}, F_{max}^{f}} \in \mathcal{R}^{1 \times H \times 1}$, denoting average and max pooling operations applied across the channel dimension on $F_{freq}$, and concatenate them. Finally, on this concatenated feature descriptor, we apply a rectangular 7x1 convolution kernel to generate a frequency attention map $M_{freq}(F^{'}) \in \mathcal{R}^{H \times 1}$, where $H$ denotes total number of frequency bins in the input feature $F^{'}$. 
\begin{equation}
    M_{freq}(F^{'}) = \sigma(f^{7\times1}([F_{avg}^{f}; F_{max}^{f}])
\end{equation}
Here, $\sigma$ denotes the sigmoid function and $f^{7\times1}$ represents a convolution operation with a rectangular $7 \times 1$ kernel.
$M_{freq}(F^{'})$ is then broadcasted along the temporal dimension on the original input feature map $F^{'}$. 


\subsection{\textit{t}-CBAM}
\label{ssec:tcbam}
\textit{t}-CBAM follows a procedure similar to \textit{f}-CBAM for modelling temporal attention, albeit limiting the receptive field of the attention module to the temporal-axis, i.e. the x-axis.

\begin{equation}
    F_{temp} = AvgPool_{H \times 1} (F^{'})
    \label{equation:avgpoolh}
\end{equation}

\begin{equation}
    M_{temp}(F^{'}) = \sigma(f^{1\times7}([F_{avg}^{t}; F_{max}^{t}])
\end{equation}

where $F_{temp} \in \mathcal{R}^{C \times H \times 1}$; and $F_{avg}^{t}, F_{max}^{t} \in \mathcal{R}^{1 \times 1 \times T}$.

\subsection{\textit{ft}-CBAM}
\label{ssec:ftcbam}

\textit{ft}-CBAM comprises of \textit{f}-CBAM and \textit{t}-CBAM applied in parallel on the input feature map. The feature maps generated by the two are then averaged.
\textit{ft}-CBAM can be seen as a special case of the original spatial CBAM, with the $7 \times 7$ convolution filter of the latter represented by two independent $7 \times 1$ and $1 \times 7$ operations.

\subsection{Proposed Pipeline}
\label{ssec:pipeline}
\textbf{CNN Front-end:} We propose a modified 50-layer Pre-Activation ResNet \cite{he2016identity}, henceforth denoted as PRN-50v2, as our CNN front-end to encode spectrogram input of arbitrary length (Table \ref{tab:cnntab}). By changing the order of layers in the residual block to BN-ReLU-Conv, pre-activation ResNets improve the ease of optimization as well as the generalization performance over comparable ResNet \cite{he2016deep} counterparts. 
\newline\textbf{Attention:} Wherever applicable, the appropriate CBAM module is integrated \textbf{at the end of each} residual block in the proposed front-end module.
\newline\textbf{Feature Aggregation:} Following \cite{xie2019utterance}, where they demonstrate the inadequacies of temporal averaging,
GhostVLAD \cite{zhong2018ghostvlad} pooling layer is applied after the CNN front-end.
For reference, experimental results using temporal average pooling are also provided.
A $256$-dimensional fully-connected embedding layer is applied after the GhostVLAD pooling layer, yielding a compact utterance-level feature descriptor.
Finally, the last fully-connected layer with softmax output for training the model in an end-to-end classification setting, using the ArcSoftmax \cite{deng2019arcface} optimization function.
\begin{table}
    \centering
    \small
    \begin{tabular}{c|c}
    \hline
    Input Spectrogram ($1 \times 161 \times T$) & Output size\\
    \hline
    conv, 7x7, 64, stride (2,1) & $64 \times 80 \times T$\\
    \hline
    maxpool, 2x2 & $64 \times 40 \times T/2$\\
    \hline
    $\Bigg[$\makecell{
        conv, 1x1, 32\\
        conv, 3x3, 32\\ 
        conv, 1x1, 64}$\Bigg]\times3$ & $64 \times 40 \times T/2$\\
    \hline
    $\Bigg[$\makecell{
        conv, 1x1, 64\\
        conv, 3x3, 64\\ 
        conv, 1x1, 128}$\Bigg]\times4$ & $128 \times 20 \times T/4$\\
    \hline
    $\Bigg[$\makecell{
        conv, 1x1, 128\\
        conv, 3x3, 128\\ 
        conv, 1x1, 256}$\Bigg]\times6$ & $256 \times 10 \times T/8$\\
    \hline
    $\Bigg[$\makecell{
        conv, 1x1, 256\\
        conv, 3x3, 256\\ 
        conv, 1x1, 512}$\Bigg]\times3$ & $512 \times 5 \times T/16$\\
    \hline
    conv, 5x1, 256 & $512 \times 1 \times T/16$\\
    \hline
    \end{tabular}
    \caption{The modified PreActResNet front-end. ReLU and BatchNorm layers are omitted. Each row depicts the filter sizes, \# of filters and the corresponding output sizes. Compared to the standard PreActResNet-50 with around 25 M parameters, the proposed model has 4.7 M.}
    \label{tab:cnntab}
\end{table}

\section{Experiments and Results}
\label{sec:exp}

\subsection{Benchmark dataset and training details}
\label{ssec:esetup}
We use VoxCeleb datasets for evaluation of the proposed approach, training our models on the VoxCeleb2 `dev' set \cite{chung2018voxceleb2} which comprises of $5,994$ speakers and test on the VoxCeleb1 \cite{nagrani2017voxceleb} verification test set \cite{chung2018voxceleb2}.
\begin{table*}[t]
    \small
    \centering
    \begin{tabular}{c|c|c|c|c|c|c}
    & Front-end Model & Front-end Attention & Dims & Aggregation & Training Set & EER (\%)\\
    \hline
    \hline
    Nagrani \textit{et al.}\cite{nagrani2017voxceleb} & I-vectors + PLDA & - & - & - & VoxCeleb1 & 8.8\\
    Nagrani \textit{et al.}\cite{nagrani2017voxceleb} & VGG-M & - & 1024 & TAP & VoxCeleb1 & 10.2\\
    Cai \textit{et al.}\cite{cai2018exploring} & ResNet-34 & - & 128 & SAP & VoxCeleb1 & 4.40\\
    Okabe \textit{et al.}\cite{okabe2018attentive} & X-vector & - & 1500 & ASP & VoxCeleb1 & 3.85\\
    Hajibabei \textit{et al.}\cite{hajibabaei2018unified} & ResNet-29 & - & 128 & TAP & VoxCeleb1 & 4.30\\
    India \textit{et al.}\cite{India2019} & CNN & - & - & MHA & VoxCeleb1 & 4\\
    Chung \textit{et al.}\cite{chung2018voxceleb2} & ResNet-50 & - & 512 & TAP & VoxCeleb2 & 4.19\\
    Xie \textit{et al.}\cite{xie2019utterance} & Thin ResNet-34 & - & 512 & GhostVLAD & VoxCeleb2 & 3.22\\
    Hajavi \textit{et al.}\cite{hajavi2019deep} & UtterIdNet & - & 512 & TDV & VoxCeleb2 & 4.26\\
    \hline
    \textbf{Proposed} & PRN-50v2 & - & 256 & TAP & VoxCeleb2 & 2.557\\
    \textbf{Proposed} & PRN-50v2 & Spatial CBAM & 256 & TAP & VoxCeleb2 & 2.515\\
    \textbf{Proposed} & PRN-50v2 & \textit{f}-CBAM & 256 & TAP & VoxCeleb2 & 2.457\\
    \textbf{Proposed} & PRN-50v2 & \textit{t}-CBAM & 256 & TAP & VoxCeleb2 & 2.28\\
    \textbf{Proposed} & PRN-50v2 & \textit{ft}-CBAM & 256 & TAP & VoxCeleb2 & \textbf{2.194}\\
    
    \hline
    \textbf{Proposed} & PRN-50v2 & - & 256 & GhostVLAD & VoxCeleb2 & 2.4\\
    \textbf{Proposed} & PRN-50v2 & Spatial CBAM & 256 & GhostVLAD & VoxCeleb2 & 2.404\\
    \textbf{Proposed} & PRN-50v2 & \textit{f}-CBAM & 256 & GhostVLAD & VoxCeleb2 & 2.13\\
    \textbf{Proposed} & PRN-50v2 & \textit{t}-CBAM & 256 & GhostVLAD & VoxCeleb2 & 2.17\\
    \textbf{Proposed} & PRN-50v2 & \textit{ft}-CBAM & 256 & GhostVLAD & VoxCeleb2 & \textbf{2.031}\\
    \end{tabular}
    \caption{Verification results on the VoxCeleb1 test set. TAP: Temporal Average Pooling, SAP: Self-Attentive Pooling, ASP: Attentive Statistics Pooling, MHA: Multi-Head Attention, TDV: Time-distributed Voting. All of the proposed models outperform existing baselines by a significant margin.}
    \label{tab:mainres}
\end{table*}

\textbf{Training Details:} For training, spectrograms are generated using a hamming window 20 ms wide with a hop length of 10 ms, and a 320-point FFT corresponding to a random 2-second temporal crop per utterance, followed by per-frequency-bin mean and variance normalization. Stochastic gradient descent optimizer with an initial learning rate of $0.01$ decayed every 15 epochs by a factor of $0.1$ is used for training.

\subsection{Experiments}
\label{ssec:exp}
Using the proposed model with no attention along with results from previous works that follow similar benchmark protocol as baselines, we first perform a direct comparative analysis to study the effect of attention on speaker verification performance. 

Further, for a more thorough assessment of the proposed attention modules and to imitate real-world conditions where similar perturbations might occur, 
we conduct three ablation experiments:
(i) random frequency masking; (ii) random temporal masking; and (iii) random frequency and temporal masking. Every input spectrogram has a 40\% probability of being augmented, with up to two mask instances per input. Per mask instance, up to 30 randomly selected frequency bins and up to 40 randomly selected timesteps are masked.

\subsection{Results}
\label{ssec:res}

Table. \ref{tab:mainres} compares the performance of the proposed models with existing benchmarks on the VoxCeleb1 test set. All the proposed models outperform previous results by a significant margin, with the best \textit{ft}-CBAM based model achieving an EER of 2.031\%. 
As evidenced by \cite{xie2019utterance}, using GhostVLAD instead of TAP improves performance all across the board. \textit{t}-CBAM variants already model temporal attention and therefore experience the smallest gains, whereas \textit{f}-CBAM variants experience the most drastic improvement (EER of 2.457 \% vs 2.13\%).

The spatial CBAM variants perform on-par with the no-attention variants of the proposed PRN-50v2 model. The large disparity in performance between spatial CBAM and \textit{ft}-CBAM can be attributed to the differences in receptive fields: unlike \textit{ft}-CBAM, the receptive field of spatial CBAM's single square 7x7 kernel will essentially span across different entities in the feature space for spectrogram inputs.
\begin{table}[h]
    \small
    \centering
    \begin{tabular}{c|c|c|c}
        Attention Variant & Temporal & Frequency & Freq+Temp\\
         \hline
         None & 2.43\% & 3.50\% & 3.54\%\\
         Spatial & 2.43\% & 3.36\% & 3.41\%\\
         \textit{f}-CBAM & 2.15\% & 3.16\% & 3.16\%\\
         \textit{t}-CBAM & 2.20\% & 3.27\% & 3.30\%\\
         \textit{ft}-CBAM & \textbf{2.05}\% & \textbf{3.01}\% & \textbf{3.05}\%\\
         \hline
    \end{tabular}
    \caption{Ablation experiment results on the VoxCeleb1 test set (EER\%). Every experiment is repeated 5 times and mean values are reported. Only GhostVLAD aggregation based models are used.}
    \label{tab:ablation}
\end{table}

Table. \ref{tab:ablation} shows the results of the ablation experiments. \textit{ft}-CBAM outperforms all other variants by a significant margin in all conditions. The performance gap between specific attention variants depends on the kind of deformation applied: the difference between \textit{f}-CBAM and \textit{t}-CBAM grows from 0.05\% (temporal masking) to 0.11\% (frequency masking).
Collectively, results from tables \ref{tab:mainres} and \ref{tab:ablation} suggest that simultaneous modelling of temporal and frequency importance improves speaker verification performance. 
\section{Conclusion}
\label{ref:conc}

In this paper, we propose methods of convolutional attention for speaker recognition, \textit{viz.} \textit{f}-CBAM and \textit{t}-CBAM for modelling frequency and temporal attention, along with a composite module that models both simultaneously, aptly named \textit{ft}-CBAM. The proposed PRN-50v2 model equipped with \textit{ft}-CBAM and GhostVLAD \cite{zhong2018ghostvlad} significantly outperforms all baselines, achieving an EER of 2.03\% on the VoxCeleb1 test set. Empirical evidence suggests that modelling attention in the DNN front-end, as well as simultaneously modelling temporal and frequency attention, improves speaker verification performance.


\newpage

\bibliographystyle{IEEEbib}
\bibliography{strings,refs}

\begin{thebibliography}{10}

\bibitem{woo2018cbam}
Sanghyun Woo, Jongchan Park, Joon-Young Lee, and In~So~Kweon,
\newblock ``Cbam: Convolutional block attention module,''
\newblock in {\em Proceedings of the European Conference on Computer Vision
  (ECCV)}, 2018, pp. 3--19.

\bibitem{nagrani2017voxceleb}
Arsha Nagrani, Joon~Son Chung, and Andrew Zisserman,
\newblock ``Voxceleb: A large-scale speaker identification dataset,''
\newblock {\em Proc. Interspeech 2017}, pp. 2616--2620, 2017.

\bibitem{chung2018voxceleb2}
Joon~Son Chung, Arsha Nagrani, and Andrew Zisserman,
\newblock ``Voxceleb2: Deep speaker recognition,''
\newblock {\em Proc. Interspeech 2018}, pp. 1086--1090, 2018.

\bibitem{Yadav2018}
Sarthak Yadav and Atul Rai,
\newblock ``Learning discriminative features for speaker identification and
  verification,''
\newblock in {\em Proc. Interspeech 2018}, 2018, pp. 2237--2241.

\bibitem{snyder2017deep}
David Snyder, Daniel Garcia-Romero, Daniel Povey, and Sanjeev Khudanpur,
\newblock ``Deep neural network embeddings for text-independent speaker
  verification,''
\newblock {\em Proc. Interspeech 2017}, pp. 999--1003, 2017.

\bibitem{hajavi2019deep}
Amirhossein Hajavi and Ali Etemad,
\newblock ``A deep neural network for short-segment speaker recognition,''
\newblock {\em Proc. Interspeech 2019}, pp. 2878--2882, 2019.

\bibitem{xie2019utterance}
Weidi Xie, Arsha Nagrani, Joon~Son Chung, and Andrew Zisserman,
\newblock ``Utterance-level aggregation for speaker recognition in the wild,''
\newblock in {\em ICASSP 2019-2019 IEEE International Conference on Acoustics,
  Speech and Signal Processing (ICASSP)}. IEEE, 2019, pp. 5791--5795.

\bibitem{zhong2018ghostvlad}
Yujie Zhong, Relja Arandjelovi{\'c}, and Andrew Zisserman,
\newblock ``Ghostvlad for set-based face recognition,''
\newblock in {\em Asian Conference on Computer Vision}. Springer, 2018, pp.
  35--50.

\bibitem{bhattacharya2017deep}
Gautam Bhattacharya, Jahangir Alam, and Patrick Kenny,
\newblock ``Deep speaker embeddings for short-duration speaker verification,''
\newblock {\em Proc. Interspeech 2017}, pp. 1517--1521, 2017.

\bibitem{Zhu2018}
Yingke Zhu, Tom Ko, David Snyder, Brian Mak, and Daniel Povey,
\newblock ``Self-attentive speaker embeddings for text-independent speaker
  verification,''
\newblock in {\em Proc. Interspeech 2018}, 2018, pp. 3573--3577.

\bibitem{okabe2018attentive}
Koji Okabe, Takafumi Koshinaka, and Koichi Shinoda,
\newblock ``Attentive statistics pooling for deep speaker embedding,''
\newblock {\em Proc. Interspeech 2018}, pp. 2252--2256, 2018.

\bibitem{India2019}
Miquel India, Pooyan Safari, and Javier Hernando,
\newblock ``{Self Multi-Head Attention for Speaker Recognition},''
\newblock in {\em Proc. Interspeech 2019}, 2019, pp. 4305--4309.

\bibitem{wang2017residual}
Fei Wang, Mengqing Jiang, Chen Qian, Shuo Yang, Cheng Li, Honggang Zhang,
  Xiaogang Wang, and Xiaoou Tang,
\newblock ``Residual attention network for image classification,''
\newblock in {\em Proceedings of the IEEE Conference on Computer Vision and
  Pattern Recognition}, 2017, pp. 3156--3164.

\bibitem{vaswani2017attention}
Ashish Vaswani, Noam Shazeer, Niki Parmar, Jakob Uszkoreit, Llion Jones,
  Aidan~N Gomez, {\L}ukasz Kaiser, and Illia Polosukhin,
\newblock ``Attention is all you need,''
\newblock in {\em Advances in neural information processing systems}, 2017, pp.
  5998--6008.

\bibitem{You2019}
Lanhua You, Wu~Guo, Li-Rong Dai, and Jun Du,
\newblock ``{Deep Neural Network Embeddings with Gating Mechanisms for
  Text-Independent Speaker Verification},''
\newblock in {\em Proc. Interspeech 2019}, 2019, pp. 1168--1172.

\bibitem{snyder2018x}
David Snyder, Daniel Garcia-Romero, Gregory Sell, Daniel Povey, and Sanjeev
  Khudanpur,
\newblock ``X-vectors: Robust dnn embeddings for speaker recognition,''
\newblock in {\em 2018 IEEE International Conference on Acoustics, Speech and
  Signal Processing (ICASSP)}. IEEE, 2018, pp. 5329--5333.

\bibitem{xu2019spatial}
Sirui Xu and Eric Fosler-Lussier,
\newblock ``Spatial and channel attention based convolutional neural networks
  for modeling noisy speech,''
\newblock in {\em ICASSP 2019-2019 IEEE International Conference on Acoustics,
  Speech and Signal Processing (ICASSP)}. IEEE, 2019, pp. 6625--6629.

\bibitem{he2016identity}
Kaiming He, Xiangyu Zhang, Shaoqing Ren, and Jian Sun,
\newblock ``Identity mappings in deep residual networks,''
\newblock in {\em European conference on computer vision}. Springer, 2016, pp.
  630--645.

\bibitem{he2016deep}
Kaiming He, Xiangyu Zhang, Shaoqing Ren, and Jian Sun,
\newblock ``Deep residual learning for image recognition,''
\newblock in {\em Proceedings of the IEEE conference on computer vision and
  pattern recognition}, 2016, pp. 770--778.

\bibitem{deng2019arcface}
Jiankang Deng, Jia Guo, Niannan Xue, and Stefanos Zafeiriou,
\newblock ``Arcface: Additive angular margin loss for deep face recognition,''
\newblock in {\em Proceedings of the IEEE Conference on Computer Vision and
  Pattern Recognition}, 2019, pp. 4690--4699.

\bibitem{cai2018exploring}
Weicheng Cai, Jinkun Chen, and Ming Li,
\newblock ``Exploring the encoding layer and loss function in end-to-end
  speaker and language recognition system,''
\newblock in {\em Proc. Odyssey 2018 The Speaker and Language Recognition
  Workshop}, 2018, pp. 74--81.

\bibitem{hajibabaei2018unified}
Mahdi Hajibabaei and Dengxin Dai,
\newblock ``Unified hypersphere embedding for speaker recognition,''
\newblock {\em arXiv preprint arXiv:1807.08312}, 2018.

\end{thebibliography}

\end{document}